# A structural and a functional aspect of stable information processing by the brain

Kaushik Majumdar

*Institute of Mathematical Sciences, Chennai – 600113, India; E-mail:* kaushik@imsc.res.in

**Abstract**

Brain is an expert in producing the same output from a particular set of inputs, even from a very noisy environment. In this paper a model of neural circuit in the brain has been proposed which is composed of cyclic sub-circuits. A big loop has been defined to be consisting of a feed forward path from the sensory neurons to the highest processing area of the brain and feed back paths from that region back up to close to the same sensory neurons. It has been mathematically shown how some smaller cycles can amplify signal. A big loop processes information by contrast and amplify principle. How a pair of presynaptic and postsynaptic neurons can be identified by an exact synchronization detection method has also been mentioned. It has been assumed that the spike train coming out of a firing neuron encodes all the information produced by it as output. It is possible to extract this information over a period of time by Fourier transforms. The Fourier coefficients arranged in a vector form will uniquely represent the neural spike train over a period of time. The information emanating out of all the neurons in a given neural circuit over a period of time can be represented by a collection of points in a multidimensional vector space. This cluster of points represents the functional or behavioral form of the neural circuit. It has been proposed that a particular cluster of vectors as the representation of a new behavior is chosen by the brain interactively with respect to the memory stored in that circuit and the amount of emotion involved. It has been proposed that in this situation a Coulomb force like expression governs the dynamics of functioning of the circuit and stability of the system is reached at the minimum of all the minima of a potential function derived from the force like expression. The calculations have been done with respect to a pseudometric defined in a multidimensional vector space.

*Keywords*: Brain circuit; Cyclic directed subgraphs; Contrast and amplify principle; Spike trains; Fast Fourier Transform (FFT); Minimization of potential function

## 1. Introduction

It's everybody's experience how effortlessly the brain recognizes a face or a voice even after a long time and in a totally unexpected situation. Significant changes in the recognizable objects might also have taken place, yet the human brain's judgment usually remains infallible. One thing is obvious, some features of the object have long been stored in the brain. As soon as the encounter happens some neurons in the brain become stimulated and a whole neural circuit spanning across different parts of the brain becomes active. As a result the stored features are recalled and a sensation of recognition dawns in the mind. If this is true, then the stored features must have been sitting somewhere in the circuit, either concentrated or scattered (latter has a greater possibility). In absence of stimulus they were just in sleep mode called *memory*, but when stimulated they woke up and have become part of the ongoing *cognition*. Again if the stimulus disappears for long the ongoing cognition will go to sleep mode as stored features i.e., will become memory. This has been termed as the *memory cognition duality* (Majumdar & Kozma, 2006).

An immediate question is "What makes the brain so robust that it can recall such an old memory in so short a time, flawlessly by being stimulated with a few stimuli even when they are corrupted with noise?" This paper will be devoted to find an answer to this question. The principal tool will be mathematical modeling.

Let us start with the notion of classical machine learning (Poggio et al., 2004). Take the training set $S = (x_i, y_i)_{i=1}^n$. Learning is finding a function $f : X \to Y$ such that $f(x_i) = y_i \forall i \in \{1,...,n\}$, where $S$ is a sample of $X \times Y$. In reality we only have $S$ and the $f$ will have to be statistically inferred (Vapnik, 1995). Obviously this inference is susceptible to error. Learning theory is concerned with *generalizability* of the learning algorithm, that is, a function $f$ will have to be chosen such a way that an upper bound of error in $f$ on $X \times Y$ can be calculated based on its error recorded on $S$. *Uniform stability* of a learning algorithm has been defined to be the phenomenon when the error bound of $f$ on $X \times Y$ becomes inversely proportional to the sample size $n$ (Bousquet & Elisseeff, 2002). This is an important notion from the Neuroscience point of view, because we know that for an average human subject too the longer is the training the better is the acquired skill and lesser is the number and amount of mistakes committed during application of that skill. However the machine learning algorithms are designed for machines with fixed architecture, whereas the brain's architecture is dynamic, for example, each time a long term memory is formed (a crucial stage of learning) brain's anatomy is altered by creating new synaptic connections (Kandel, 2006). To cope with the complexity of brain and our lack of precise knowledge about how it works the notion of stability in neuroscience has evolved in a different and less rigorous manner. A good example of stability in neuroscience is the identification of animal hippocampus place cells with the long term memory of spatial locations (for a review see Kentros, 2006). Similar observation for human hippocampus has also been reported (Ekstrom, et al., 2003). Intuitively *stability of information processing* by the brain should mean the ability to extract the right output from the input when some essential features (identified in its own peculiar way by an individual brain) are present, possibly with significant amount of noise, in that input. For a brain the noise may not always be totally irrelevant information, but information with substantial potentiality to be a valid output. Ability to choose a particular one or at most a few ones from among a large number of alternatives over and over again and again is also a remarkable example of stability shown by the brain.

Activity dependent modulation of synaptic strength and structure is emerging as one of the key mechanisms by which information is processed and stored within the brain (Bailey & Kandel, 2004). This points towards a model of brain architecture by directed graphs, where each node is a neuron and edges are synaptic connections (Majumdar & Kozma, 2006). In section 2 a new principle named as *contrast and amplify* principle will be described, by which brain is likely to separate relevant information from the irrelevant ones and amplify them in course of propagation through the networks. In section 3 information extraction from neuronal spike trains by fast Fourier transform will be described. It is hypothesized that absolute minimization of a potential function with respect to a pseudometric in this information space helps the brain to choose from among different alternatives. The two methods described in section 2 (from a structural point of view) and section 3 (from a functional point of view) respectively seem to be important in maintaining stability of information processing by the brain. It should be mentioned here that both the methods are biologically plausible. In this paper they have been treated together from a more physical (and less biological) point of view to emphasize the aspects of stability in learning and memorizing by the brain. Each of them will be taken up separately in two different papers (Majumdar, 2007a and Majumdar, 2007b for the functional and structural parts respectively with neurobiological applications) and will be elaborated on their biological viability and applicability.

## 2. Structural stability

The closest computational analog of a neuronal circuit in the cortex (where most of the computations in a mammalian brain are carried out) is a directed multigraph, whose each node will represent a neuron and each edge a synapse. It can be represented as a three dimensional array $a[i, j, k]$. If there are $p$ different synapses joining the neuron $i$ with the neuron $j$ (assuming that all neurons in the brain have been numbered) then $a[i, j, k]$ will give the weight (the gain in synaptic transmission) of the $kth$ synapse joining the neuron $i$ with the neuron $j$, where $1 \leq k \leq p$. If the neuron $i$ and neuron $j$ are excitatory then $a[i, j, k]$ will be positive, if they are inhibitory then $a[i, j, k]$ will be negative. For given values of $i, j$ and $k$ $[i, j, k]$ uniquely determines a synapse and $a[i, j, k]$ represents the synaptic weight. In general $a[i, j, k]$ will be a function of time and may be written as $a[i, j, k](t)$ or $a_{ijk}(t)$, where $t$ denotes time.

In artificial neural networks (ANN) synaptic weights are well defined and they are fixed during training or learning. Then they continue to process data with fixed valued synaptic weights. On the other hand in the neural network of the brain a single synapse may display several varieties of synaptic enhancements (Castro-Alamancos, et al., 1996). So $a_{ijk}(t)$ will be an interesting function to study. But it is extremely difficult to measure. The postsynaptic and presynaptic cells of an excitatory synapse should be in phase synchronization. A strict measure of phase synchronization may be a good measure of $a_{ijk}(t)$ in this case. Phase synchronization can be measured either statistically (Tass, et al., 1998 in not so strict sense) or deterministically (Majumdar, 2006a in a very strict sense). For both the methods neuronal spike trains will have to be treated as ordinary signals and two neurons are firing synchronously if and only if their spike trains are in phase synchronization. The deterministic method is likely to give a better result through the measure of synchronization, the *syn* function, defined in (Majumdar, 2006a & 2006b). In fact if zero time lag phase synchronization is detected between two neurons (Majumdar, 2006a) there are reasons to believe they are strongly coupled by excitatory synapses. If a pair of neurons are firing asynchronously they may either be connected by inhibitory synapses or they may not have a common synapse at all. However if they are quite close to each other chances are higher that they are coupled with inhibitory synapse(s). In this case also the value of *syn* function will be able to measure the degree of asynchrony between the two neurons.

As soon as the brain receives information from the environment through the sensory excitatory neurons a sub-circuit of the whole cortical circuit is activated to start the cortical computation in order to process those information, which culminates in some form of cognition (with or without a motor action in general). Let us denote all active receptor neurons as $T_0$ or target 0 neurons. The neurons fired by $T_0$ be called $T_1$ neurons or target 1 neurons. In general $T_i$ be the class of neurons fired by the class of $T_{i-1}$ neurons. These classes are not mutually exclusive. Some forward class neuron can excite back an earlier class neuron and in fact this happens quite often. Several retrograde messengers have been identified that once released from dendrites act on presynaptic terminals to regulate release of neurotransmitters (Abott & Regehr, 2004).

**Definition 2.1:** A *big loop* will mean a cyclic brain circuit consisting of a longest feed forward path from the sensory neurons to the highest processing brain region and a feed back path from that region to closest to the same sensory neurons (Figure 2.1). In other words it is a chain of $T_0,....,T_n$ neurons.

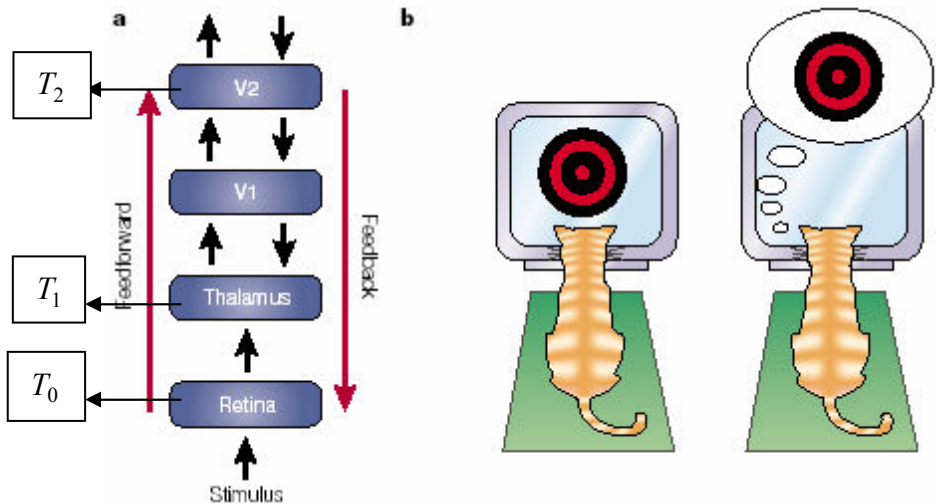

Figure 2.1 A big loop consisting of stimulus driven feed forward and expectation driven feed back paths (a). The stimulus and the expectation for the stimulus in the brain of the animal has been illustrated in (b). The information processing stages have been mentioned by 0th order, 1st order, etc. Adapted from Engel, et al. (2001) and reproduced with permission from the Nature Publishing Group.

Notice that the big loops are traversing through different functional regions of the brain and during traversal through the big loops signal or information about the sensory input gets processed hierarchically. The notion of information processing by feed forward path (the so called bottom up approach) is classical but the notion of processing by feed back path (top down approach) is relatively modern (see Engel, et al., 2001 for a review). The notion has been graphically illustrated in Figure 2.1.

Not only inhibitory interneurons are activated by feed back path, but also the same happens for excitatory interneurons. The top down signal carries message through both excitatory and inhibitory interneurons, and excitatory and inhibitory synapses.

**Lemma 2.1:** A cycle with $k$ nodes (neurons) in the directed graph of the brain circuit can be activated with greater probability than a linear path or line with the same number of nodes.

**Proof:** In a directed cycle if any neuron is activated the signal will propagate to all the other nodes cyclically and they will become activated in turn. Whereas to activate all the nodes in a line the first node must have to be activated. So if any neuron in the cycle can be activated with probability $p$ the whole cycle will be activated with probability $p$, whereas the whole line will have activation probability $\frac{p}{k}$ only. □

**Lemma 2.2:** Cycles (in the sense of directed graphs or digraphs) in brain circuit may amplify a signal.

**Proof:** Let a signal or an action potential of frequency $I$ (in case of a firing neuron frequency signifies intensity, so $I$ is also intensity of the signal) is reaching the $jth$ node of a cycle $(n_1,...,n_j,...,n_k)$, where $n_{k+1} = n_1$. If a signal takes time $T$ on an average to travel from one node to the next, and the incoming signal to $n_j$ is still reaching the node after $(k-1)T$ time, then the total dendritic input to $n_j$ becomes $I + I'$, where $I'$ has been received through the feed back loop. If $n_j$ is not already firing at the highest frequency the input $I + I'$ will make it fire at higher frequency than did $I$. □

However automatic amplification blow up according to Lemma 2.2 does not happen, for natural constraint on the firing frequency of a neuron. Note that by the method of Lemma 2.2 it is possible for the brain to amplify an internally generated stimulus leading to subsequent activation of a larger circuit to give rise to internally generated cognitive processes. The whole thing can happen without the presence of an external stimulus.

**Claim 2.1:** An activity pattern consisting of feed forward paths (FFP) emerges based on memory of past experience.

**Answer:** As an FFP reaches a higher processing region like the hippocampal formation or the prefrontal cortex it gets access to neurons with more diverse connections and greater processing power. Let $A$ and $B$ be two FFPs as collections of feed forward lines (a feed forward line is a chain of neurons connected by excitatory feed forward synapses and a collection of such closely spaced lines is an FFP). $u$ be a neuron in the hippocampal formation connected to all feed forward lines belonging to $A$ (presynaptically by dendritic arbor) and $B$ (postsynaptically by axonal branching) which will fire if at least $k$ presynaptic neurons fire. Now if sensory information coming through thalamus activates $k$ or more feed forward lines of $A$ then the probability that $B$ will also be activated is $\sum_{j=k}^{r} \frac{h!}{j!(h-j)!} p_0^j (1-p_0)^{r-j}$, where $p_0$ is the probability of firing a presynaptic neuron to $u$ and $h$ is the number of activated feed forward lines in $A$. To increase the chance of activating $B$ several $u$ will be needed. The higher is the number of $u$ the greater will be the success of activating $B$ through $A$. Past experience will determine the exact value of $p_0$ (signifying synaptic potentiation) and $u$ (signifying synaptic connections formed). Clearly $p_0$ and $u$ together represent memory of past experiences (for elaboration of the last sentence see Kandel, 2001). □

**Theorem 2.1** (*Contrast and amplify* principle)**:** During the bottom up journey signal travels from sensory neurons to the highest processing brain areas by contrast and amplify principle.

**Proof:** Sensory inputs are carried by parallel feed forward lines in the graph of the brain circuit from the sensory neurons to the highest processing brain areas. Any two of the parallel lines may be short circuited (i.e., joined together) by excitatory and/or inhibitory synapses. At the very beginning all the parallel lines carry signals from the $T_0$ neurons according to the collection of FFPs available at that instant, i.e., all the lines are active and between any pair of them excitatory and inhibitory synapses are active or inactive without any control from the higher order brain regions (say $T_n$ neurons). Once the signals reach in those regions an activity pattern emerges based on memory of past experience (Claim 2.1), which is the *expected* pattern (as described in Engel et al., 2001) and it in turn starts controlling the $T_j$ (for some $j's$) neurons of the FFPs through feed back lines, where $0 < j < n$. A bunch of closely spaced feed forward lines short circuited by excitatory synapses through various $T_j$ neurons work as a single path. Two such groups of paths short circuited mostly by inhibitory interneurons work as different paths. This way some information which are flowing through the same path are processed together as a cluster and two clusters become distinguished, because they are information carried by different paths. In the process the whole sensory input is decomposed by contrast.

Next, take a path as a collection of parallel lines in the directed graph of the brain circuit, which are pair wise short circuited mostly by excitatory synapses. These short circuiting excitatory synapses when form (directed) cycles can enhance the signal passing through it according to Lemma 2.2. By the argument of Lemma 1.1 the signal enhancement will be more if $T$ is small, such as if most of the edges of the cycle are electrical synapses, $k$ is small and $I'$ is large (ideally close to $I$). □

By now the structure of a big loop is somewhat clear. Information processing by a big loop separates out important features from a noisy input and then amplify them to create a strong representation of the object of interest. This way the stability of information processing during identification of an object is preserved.

## 3. Functional stability

Although a neural circuit is a dynamic structure, for a short period of time (of the order of minutes) it remains structurally unchanged. However even within this short period the function of the network is highly dynamic. Despite the complex physiological processes going on in the network the outputs are produced in the form of action potentials. Many behavioral tasks (such as gill withdrawal of *Aplysia* in response to a tactile stimulus (Kandel, 1976) or knee jerk by humans in response to a tapping (Kandel, et al., 2000)) can be accounted for the action potentials generated by the neurons in the network. The behavior of the network may be thought of as an assembly of behaving (firing) neurons. Behavior of each neuron can be observed in the train of action potentials (the so called spike train) of the neuron. Now the question is how to record the information emanating out of all the neurons simultaneously in a neural network? First it will be shown a neuronal spike train can be expressed as a Fourier series.

One of the earliest Fourier analysis on neuronal spike trains was carried out in (Lange and Hartline, 1979). Fourier transform or Fourier spectrum decomposition of neuronal spike trains has been reported in (Schild, 1982 and Reich, et al., 1997). Usually a neuronal spike train is sampled either as a continuous signal or a point process or as a series of Dirac delta functions. Advantages and limitations of each of them have been discussed in (French & Holdon, 1971). It is vital to retain as much information as possible about the shape of a spike train (such as sculpturing (Fig. 3.1)) and therefore a piecewise continuous function representation of the spike train over the interval of interest is preferred.

Let $f$ be a periodic function, with period $p$, defined on the closed bounded interval $[0, p]$. Then

$$f(t) = \frac{a_0}{2} + \sum_{n=1}^{\infty} (a_n \cos \frac{2\pi n t}{p} + b_n \sin \frac{2\pi n t}{p}) \tag{3.1}$$

for $t \in [0, p]$. $f$ can be extended to the whole real line by dividing the line into closed and bounded intervals of length $p$ each. $a_n$ and $b_n$ are given by

$$a_n = \frac{2}{p} \int_0^p f(t) \cos \frac{2\pi n t}{p} dt \tag{3.2}$$

$$b_n = \frac{2}{p} \int_0^p f(t) \sin \frac{2\pi n t}{p} dt. \tag{3.3}$$

The right hand side of (3.1) will converge to the left hand side according to the following theorem.

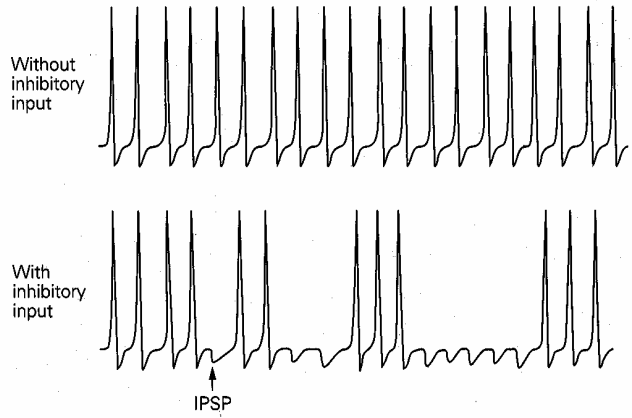

Figure. 3.1. A neuron usually receives both excitatory and inhibitory inputs simultaneously. Without inhibitory input the neuron fires continuously at a fixed interval. With inhibitory input some action potentials are inhibited, resulting in a distinctive pattern of impulses. The effect of inhibition on the firing of a neuron is called *sculpturing*. IPSP stands for inhibitory post synaptic potential. Adopted from (Kandel, et al., 2000).

**Theorem 3.1:** Suppose that $f$ has period $p$, and suppose that $t_0$ is a point in $[0, p]$ where $f$ has one-sided limiting values and one-sided derivatives. Then the Fourier series of $f$ converges for $t = t_0$ to the mean value $\frac{1}{2}(f(t_0+) + f(t_0-))$. In particular, if $f$ is continuous at $t_0$, the sum of the Fourier series equals $f(t_0)$.

**Proof:** See (Vretbald, 2003) (Proof of Theorem 4.5, where the period has been taken as $2\pi$, but the same proof is valid for any other period). □

Perhaps the most faithful visualization of action potentials and spike trains can be accomplished through an oscilloscope. A spike train can reasonably be taken as a graph of a piecewise continuous, bounded function. It is also clear from the oscilloscope pictures that the function has left and right derivatives at the points of discontinuity. The spike train is a wavy signal where waves are accentuated very sharply to form spikes. So by Theorem 3.1 a Fourier series representation of a spike train is valid. Hence the underlying Fourier series of a neuronal spike train can be recovered by Fourier transform. The Fourier transform (or rather the fast Fourier transform or FFT) produces the vector $a_0 a_1 b_1 a_2 b_2 .... a_r b_r$. For convenience it can be rewritten as $e(1).....e(2r+1)$, where

$$\left.\begin{aligned} e(1) &= a_0 \\ e(n) &= a_{n/2} \quad \text{if } n \text{ is even} \\ e(n) &= b_{2n-1} \quad \text{if } n \text{ is odd} \end{aligned}\right\}, \qquad (3.4)$$

in uniform symbol. More conventionally the vector can be written as $(e(1),...,e(2r+1))^T$. The suffix $T$ stands for transpose. For good result the sample frequency for FFT should be 1000 Hz. If the period is $p$ seconds then $r = 500p$. Now if there are $N$ neurons in the behaving network under study, there will be $N$ number of points in the $2r+1$ dimensional vector space for a behavioral study spanning $p$ seconds. Let the *kth* vector be denoted as $(e_k(1),...,e_k(2r+1))^T$, where $k \in \{1,....,N\}$. Notice that each of the $N$ points thus constructed in the $1000p+1$ dimensional vector space represents information contained in a neuronal spike train up to the sculpturing as shown in Figure 3.1. Also note that since the Fourier series is convergent $e_k(2r+1) \to 0$ as $r \to \infty$. This means after some time the behavior does not change significantly unless new neurons are recruited or neurotransmitter release is significantly altered in the network. Both of these will need new internal or external stimuli.

So far the information extraction from the neuronal spike trains has been discussed to be by Fourier transform. This is only one of several possible methods. There is no single method which works equally well for all types of spike trains, as for example, Fourier transform works best when there is a tonic bursting of spikes. On the other hand when the spikes are very few and far between in a spike train the FFT may have to be applied more carefully. For a good result the volume of data must have to be quite substantial. One way to do this is to improve the sample frequency. Since duration of a single spike is from 1 to 2 ms the sample frequency must be 1000 Hz or more, otherwise there may be several spikes undetected. In fact increasing the sample frequency is a fundamental requirement irrespective of the processing technique adopted. The advantage of FFT is that when the sample frequency and the signal to noise ratio (SNR) are sufficiently high it picks up very minute details like the sculpturing (Figure. 3.1) in a spike train.

**Definition 3.1:** The cluster of $N$ points $(e_k(1),...,e_k(2r+1))^T$, where $k \in \{1,....,N\}$ gives the *behavioral map* of the network for duration $p$.

$r = \dfrac{f}{2}p$, where $f$ is the sample frequency. Although it is a standard practice among the computational neuroscientists to represent the behavior of a neuronal network by neuronal spikes alone, it is important to note that even when a neuron is oscillating below threshold for spike initiation, it can still release neurotransmitter and shape the final circuit output (Harris-Warrick & Marder 1991). But for simplicity of modeling I shall ignore this fact in this paper.

When neurons are very closely spaced they fire synchronously with greater probability (Maldonado et al., 2000). A cluster of closely spaced synchronously firing neurons may be taken as a single neuron. Suppose $B^1$ and $B^2$ are two different behaviors. The maps represent the behaviors are $(e_k^1(1),...,e_k^1(2r+1))^T$ and $(e_k^2(1),...,e_k^2(2r+1))^T$ respectively for $k \in \{1,...,N\}$.

**Definition 3.2:** Distance between two behaviors or the behavior maps is defined as

$$d(B^1, B^2) = \frac{1}{N} \sqrt{\sum_{i=1}^{2r+1} \left[ \sum_{k=1}^{N} \{e_k^1(i) - e_k^2(i)\} \right]^2} . \tag{3.5}$$

$d$ will be called *cluster distance* between two behaviors of the same network.

Note that $d$ is not a metric, for $d(B^1, B^2) = 0$ does not necessarily mean $B^1 = B^2$. To verify this consider a two neuron network, where in $B^2$ the first neuron has the spike train identical to the spike train of the second neuron during $B^1$ and the second neuron during $B^2$ has spike train identical to the first neuron during $B^1$. Clearly $d(B^1, B^2) = 0$, but $B^1 \neq B^2$. For $B^1, B^2$ and $B^3$,

$$d(B^2, B^3) = \frac{1}{N} \sqrt{\sum_{i=1}^{2r+1} \left[ \sum_{k=1}^{N} \{e_k^2(i) - e_k^3(i)\} \right]^2} , \tag{3.6}$$

and

$$d(B^1, B^3) = \frac{1}{N} \sqrt{\sum_{i=1}^{2r+1} \left[ \sum_{k=1}^{N} \{e_k^1(i) - e_k^3(i)\} \right]^2} . \tag{3.7}$$

Write $\sum_{k=1}^{N} e_k^1(i) = C(i)$, $\sum_{k=1}^{N} e_k^2 = D(i)$ and $\sum_{k=1}^{N} e_k^3 = E(i)$ in (3.5), (3.6) and (3.7). Since $\{C(i)\}_{i=1}^{2r+1}$, $\{D(i)\}_{i=1}^{2r+1}$ and $\{E(i)\}_{i=1}^{2r+1}$ are all vectors in the Euclidean space $R^{2r+1}$ ($R$ stands for set of real numbers), their mutual Euclidean distances satisfy triangle inequality. This implies $d(B^1, B^3) \leq d(B^1, B^2) + d(B^2, B^3)$. So the following result has been established.

**Theorem 3.2:** Cluster distance is a pseudometric.

The usual definition of derivative for the real numbers can be extended to the pseudometric space $(R^{2r+1}, d)$, which is the space of behavioral map. In notion of stability in the space of behavioral map differentiability will be important. $N$ plays only a superficial role in all the computations. With current recording techniques $N$ can at best be a hundred or so. However if by some inferential, or other method data from a large number of neurons can be incorporated then $N$ will be a significant quantity and to keep the value of $d$ away from blowing up due to large $N$ the current form of the distance function, where an 'averaging' is performed, will be necessary.

Let in a given network memory of $m$ number of behaviors has already been stored. Now in response to environmental stimuli the network has been activated in order to generate a new behavior map. This map will have to be generated in a 'stable' manner. In the last section stability meant the ability of the processor to produce the 'right' output even when the important features of that object in the input signal are buried under noise to some extent. In this section the meaning of stability will be slightly more sophisticated. Here the noise may consist important, interesting and meaningful information, like the situation where a choice is to be made out of several potential options.

**Principle 3.1** (hypothetical)**:** In the memory of $m$ behaviors in a network let $s$ be the number of behaviors with which purely positive experience is associated and with the remaining $m-s$ purely negative experience is associated. The new behavior in response to stimuli by activating the same network will be generated in such a way that its map in $\left(R^{2r+1}, d\right)$ will be as far away from the maps of the $m-s$ behavior maps and as close to the $s$ behavior maps as possible.

Let me elaborate. Positive experience and negative experience are purely subjective. One way of describing a positive experience of an individual may be an experience whose reoccurrence is desired by that individual for that moment. Similarly a negative experience of an individual may be the one whose reoccurrence is undesirable to that individual for that moment. There are behaviors with both positive and negative experiences associated with them. Those behaviors will be called *complex behavior* and they can be decomposed into simple behaviors where each *simple behavior* will only have either purely positive or purely negative or purely neutral (i.e., neither positive nor negative) experience associated with it.

The words 'far away' and 'close to' are only metaphorical in the Principle 3.1. However it is clear that some kind of optimization is necessary. Let us consider an analogous situation in physics. Let there be seven charged particles on a plane and they are all in arbitrary but fixed positions. Four of them are positively charged and three are negative. A new negative charge is introduced from outside a bounded region of the plane which includes all the fixed positioned charges. Now at which points within the bounded region the potential energy requirement to place the new charge will be minimum?

Let the locus of the introduced negative charge is $(x, y)$. The position of four positive charges be $\{(x_i, y_i)\}_{i=1}^{4}$ and that of the three negative charges be $\{(x_i, y_i)\}_{i=5}^{7}$. The Coulomb force $F(x, y)$ acting on the free charge is given by

$$F(x,y) = -\sum_{i=1}^{4} \frac{C}{(x-x_i)^2 + (y-y_i)^2} + \sum_{i=5}^{7} \frac{C}{(x-x_i)^2 + (y-y_i)^2}, \qquad (3.7)$$

where $C$ is a constant.

Let the new behavior of the net mentioned in Principle 3.1 be $B$. $B$ will be attracted towards $s$ behaviors with positive experience and will be repulsed by the remaining $m-s$ behaviors with negative experience. In this sense $B$ will try to be as 'far away' from the set of $s$ behaviors and as 'close to' the set of $m-s$ behaviors as possible. By analogy with (3.7) I can write the governing expression of the dynamics as

$$G(B) = -\sum_{i=1}^{s} \frac{P_i(X)w_i(t)}{(d(B_i,B))^2} + \sum_{i=m-s+1}^{m} \frac{P_i(X)w_i(t)}{(d(B_i,B))^2}, \quad (3.8)$$

where $G$ stands for *governing expression*, $P_i(X)$ is a multivariate independent normal probability distribution function with mean $B_i$ and a variance for each component, $w_i(t)$ is the *interaction weight* between $B_i$ and $B$. $P_i(X)$ signifies emotion. To be more precise for a behavior with positive experience it gives the measure of how positive the experience is. Similarly for a negative experience it gives the measure of how negative the experience is. In case of neutral experience $P_i(X)$ becomes Dirac's delta function, which is unity for $X = B_i$ and zero otherwise. $w_i(t)$ will generally depend on $a_{luv}(t)$, where $a_{luv}(t)$ varies over all the synapses in the net. The whole issues of long term potentiation (LTP), short term potentiation (STP) and long term depression (LTD) will come into play. However for a short period of time, spanning up to a few minutes $w_i(t)$ may be taken as constant and $w_i(t) = w_j(t)$, for all $i, j \in \{1,...,m\}$.

Unlike in the charged particle system described above $P_i(X)$ plays a very important role in the behavior dynamics. Intuitively $P_i(X)$ keeps the effect of $B_i$ or the range of 'force' due to $B_i$ small depending upon the value of variance for each component of $X$. The eigenvalues of the covariance matrix of $P_i(X)$ will be small if the emotion involved with the behavior $B_i$ is low. Likewise the eigenvalues of the covariance matrix of $P_i(X)$ will be large if the emotion involved with the behavior $B_i$ is high.

The system governed by $G(B)$ is a holonomic system (Goldstein, 1950), where each point of $B$ will be on a hypersurface in $(R^{2r+1}, d)$. $B$ will act like a rigid body. If $G$ is treated like a force from (3.8) it is clear that $G$ can be derived from a scalar valued function $V(G)$ such that

$$V(B) = \sum_{i=1}^{s} \frac{P_i(X)w_i(t)}{d(B_i,B)} - \sum_{i=m-s+1}^{m} \frac{P_i(X)w_i(t)}{d(B_i,B)}. \quad (3.9)$$

Since for a short time $w_i(t)$ are constant for all $i$, $V(B)$ is time independent and the system is called conservative. At that value of $B$ where $V(B)$ attains the absolute minimum, i.e., minimum of all the minima, the system will be stable. In that position some perturbation in the value of $B$ will not alter the system permanently, but will bring it back to the absolute minimum $B$ position.

Let us now consider a simple example. Take the knee jerk behavior of a human in response to a tapping (Kandel et al., 2000). The foreleg is dragged behind as a reflex action in response to a sudden tapping in front. Now if a sharp edge is placed behind the foot at a little distance so that the foot hits the edge during backward motion then after one experience (or may be a few times more in case of a not so cautious subject) the subject will still drag the foot backward in response to a tapping, but to a lesser extent in order to avoid hitting the sharp edge on the other side. Being tapped is a (passive) behavior with negative experience, say $B_1$. Also being hit by a sharp edge is a behavior with negative experience, say $B_2$. The potential function is

$$V(B) = -\frac{P_1(X)w_1(t)}{d(B_1,B)} - \frac{P_2(X)w_2(t)}{d(B_2,B)}. \tag{3.10}$$

According to the hypothesis $B$ (the behavior which pulls back the foreleg from the tapping, yet does not come too backwards to hit the sharp edge) will have to be such that $V(B)$ attains the smallest minimum. Here $B$ attains only in the denominator and $w_1(t) = w_2(t) = const$.

$$d(B_1, B) = \frac{1}{N}\sqrt{\sum_{i=1}^{2r+1}\left[\sum_{k=1}^{N}\left\{e_k^{B_1}(i) - e_k^{B}(i)\right\}\right]^2}. \tag{3.11}$$

$$d(B_2, B) = \frac{1}{N}\sqrt{\sum_{i=1}^{2r+1}\left[\sum_{k=1}^{N}\left\{e_k^{B_2}(i) - e_k^{B}(i)\right\}\right]^2}. \tag{3.12}$$

For minimization of $V(B)$ the following must hold

$$\frac{\partial V(B)}{\partial(i,k)} = 0, \quad \forall i \in \{1,....,2r+1\}, \quad \forall k \in \{1,.....,N\}, \tag{3.13}$$

where $\partial(i,k) = \partial e_k^B(i)$. Each of the $N(2r+1)$ equations in (3.13) are nonlinear and therefore may have to be solved numerically to determine $B$ ($B$ has altogether $N(2r+1)$ number of unknowns). A more detailed derivation of this solution has appeared in (Majumdar, 2007a), where a few neurobiological applications of the derivation have also been discussed.

Intuitively the minimization of $V(B)$ cannot occur close to either $B_1$ or $B_2$, for in those neighborhoods $G(B)$ will have a high value. However because of $P_1(X)$ that value will diminish exponentially fast as $B$ moves away from $B_1$. Similarly the case for $B_2$ due to $P_2(X)$. This means minimization of $V(B)$ will occur for a $B$ spaced apart from both $B_1$ and $B_2$. This intuitively makes sense for the knee jerk behavior described above, for the foreleg will position itself some where in between the positions of the two negative behavioral experiences.

Actually if we apply more physical insight into (3.13) some reduction is possible, by which the value of $N$ will go down and with this the value of $N(2r+1)$ will also be reduced. Lesser number of equations in (3.13) will have to be dealt with. In fact in case of $B_1$ and $B_2$ the neural circuit will behave almost in the same manner, that is, the neurons will generate spike trains in almost the same manner except for some of the sensory neurons in the front part of the foreleg (where tapping is made) and some of the sensory neurons in the back (where the leg is hit by the sharp edge) and the motor neurons controlling dragging and moving forward of the foreleg. For all practical purposes the study of $B_1$, $B_2$ and $B$ can be kept confined in those neurons only. The duration of the behaviors is only a couple of seconds and therefore $r$ will not be high at a sample frequency 1000 Hz or even more.

**Discussion and conclusion**

In this paper stability in information processing by an individual brain has been analyzed from structural and functional point of view. It has been shown that forming loops by feed forward and feed back paths during cortical rewiring to preserve long term memory can facilitate stable recollection. On the other hand even when the rewiring is conducive to stable recollection the neurons of that rewired network need to follow certain activity pattern for stable recollection. For this a potential function has been introduced and minimization of that function has been identified as a sufficient condition for stable information processing leading to stable recollection of past experiences. Stability of information processing here means the remarkable ability of a brain to reproduce the same behavior whenever some particular features are present in the stimuli from the environment even along with significant distracting noise. A good example is a familiar face recognition in a very unexpected environment. It has been assumed here that the collective behavior of the neurons present in the circuit during a particular task makes up the functional behavior of that circuit. This may be a naive assumption (apart from firing of the neurons there are other activities in the brain which account for its over all performance), but it has been shown that this assumption gives an intuitively appealing mathematical model. According to this model for a behavior $B$ to be stable the absolute minimum or minimum of all the minima of the potential function $V(B)$ in (3.9) will have to be reached (this has been further explored in (Majumdar, 2007a)). This requirement is in addition to the architectural requirement, that is, even when the neural circuit contains the appropriate loops for the stable processing of information (described in section 2). (3.9) comes from a more fundamental expression given by (3.8). Note that (3.8) contains terms for synaptic plasticity $w_i(t)'s$ and memory $B_i's$. When the time is long enough so that $w_i(t)$ changes significantly (3.9) cannot have that simple form. In this paper the form of $B_i$ has been made some what clear, but unfortunately this has not been true about $P_i(X)$ and $w_i(t)$. This will remain open for future exploration.

(3.8) may be one of several forms in which behavior dynamics of a neural circuit can be represented or modeled. Like the Fourier analysis the neuronal spike trains can also be subjected to wavelet analysis. That will give different representation of the behavior map. Still the fundamental governing equation of the behavior dynamics will retain in the form of (3.8), only except Fourier coefficients substituted by wavelet coefficients. Since (3.8) gives the behavior dynamics involving both memory and synaptic plasticity it will be worth exploring what form $V(B)$ should take when $w_i(t)'s$ are changing.

If greater neuronal activity requires greater blood flow, greater metabolism and greater sugar intake then $B_i's$ should have specific fMRI (functional Magnetic Resonance

Imaging) and PET (Positron Emission Tomography) imprints. So also does $B$. Next we need to know at least an empirical relationship between metabolic rate and action potential generation in case of a single neuron. Once that is known the fMRI and PET imprints of the behaviors can be translated into action potential or spike train imprints. FFT of those imprints will give vector representation of behaviors as described in section 3. How the images of $B_i's$ are interacting with that of $B$ will give ways to estimate $P_i(X)'s$ (since they are multivariate independent normal distribution with known mean only the variances are to be estimated, which will give a numerical measure of emotion). It may also be possible to estimate $w_i(t)'s$ this way. So $V(B)$ can be estimated through fMRI and/or PET studies. In turn $B$ can be predicted if good estimates for $B_i's$ and $w_i(t)$ are available. Some progress has been made in these directions in a subsequent paper (Majumdar, 2007a).

## Acknowledgement

The author likes to acknowledge the Institute of Mathematical Sciences for a postdoctoral fellowship under which this work has been carried out. Helpful comments by three anonymous reviewers are also being acknowledged. One of them pointed out the works of E. R. Caianiello and coworkers with which my developments may have some similarities. Unfortunately I could not access any material on their works at the time of reviewing this paper.